# Strong magnetic fields in normal galaxies at high redshifts*

Martin L. Bernet[1], Francesco Miniati[1], Simon J. Lilly[1], Philipp P. Kronberg[2,3], Miroslava Dessauges-Zavadsky [4]

[1]Department of Physics, ETH Zürich, Wolfgang-Pauli-Strasse 16, CH-8093 Zürich, Switzerland.

[2]Los Alamos National Laboratory, IGPP, P.O. Box 1663, Los Alamos NM 87545, USA.

[3]Department of Physics, University of Toronto, 60 St. George St., Toronto M5S 1A7, Canada

[4]Observatoire de Genève, 51 Ch. Des Mailletes, CH-1290 Sauverny, Switzerland

**The origin and growth of magnetic fields in galaxies is still something of an enigma[1]. It is generally assumed that seed fields are amplified over time through the dynamo effect[2-5], but there are few constraints on the timescale. It has recently been demonstrated that field strengths as traced by rotation measures of distant quasars are comparable to those seen today[6], but it was unclear whether the high fields were in the exotic environments of the quasars themselves or distributed along the line of sight. Here we demonstrate that the quasars with strong MgII absorption lines are unambiguously associated with larger rotation measures. Since MgII absorption occurs in the haloes of normal galaxies[7-11] along the sightline to the quasars, this association requires that organized fields of surprisingly high strength are associated with normal galaxies when the Universe was only about one-third of its present age.**

In a recent study[6,12] we used a large sample of extragalactic radio sources to investigate the redshift evolution of the Faraday Rotation Measure (RM) of polarized quasars up to $z \sim 3$. As had been suggested by some earlier studies with more limited data[13-16], we found that the dispersion in the RM distribution of quasars increases at higher redshifts. This is a non-trivial finding due to the strong $(1+z)^{-2}$ dilution effect, which works in the opposite direction. The physical interpretation of this result remained however uncertain: in particular, there was no way of knowing whether the increase in RM observed at higher redshifts was associated with an evolution of the exotic and unusual environs of the very luminous radio quasars (an "intrinsic" origin), or with the presence of magnetic fields in more typical environments distributed along the sightlines to these cosmologically distant sources (an "intervening" origin).

In order to test between these possibilities and thereby resolve a long-standing debate[13-16] about the cause of redshift evolution of the RM in a clear and decisive way, we have now obtained high resolution spectra for 76 quasars, selected from our full RM sample[6]. The spectra (available online as 'Supplementary Information') were taken using the UVES[17] spectrograph on the ESO 8-m Very Large Telescope (VLT). This enabled us to conduct a census of intervening material along each individual sightline. The quasars were selected for spectroscopy by requiring that their redshifts lie in the range $0.6 < z < 2.0$, spanning the range where the RM distribution is seen to broaden[6]. Only quasars at Galactic latitudes $|b| > 30°$ were observed, to minimize problems of Galactic foreground RM contamination. To ensure adequate signal to noise ratios in the optical spectra, the quasars were all selected to be brighter than $m_v \sim 19$. No selection relative to the observed RM of the quasars was made.

MgII absorption systems in the optical spectra were identified through the double absorption feature at rest frame wavelengths 2796.35 and 2803.53 Å. We catalogued MgII absorption systems with rest-frame equivalent width greater than $EW_0 > 0.3$ Å in the 2796 Å line where our sample should be complete. Because of reduced signal to noise ratios in the ultraviolet, we discarded the spectra below 3760 Å and thus any MgII systems lying below redshift $z =$

0.35 (few would be expected). In order to preserve the objectivity of the experiment, the detection of MgII systems was carried out completely blind to the RM of the quasar.

To provide the cleanest possible test of the intervenor hypothesis, we removed two quasars from the sample that exhibit MgII absorption at the same redshift as the quasar itself, and also discarded three other quasars in which the optical and radio sightlines are separated by more than 7 arcsec ( ¿ 50 kpc at the redshift of the MgII absorption). However, there is no significant change in our results if these deletions are not made. Our final sample thus consists of 71 quasar spectra which together cover a total redshift path interval of $\Delta z = 52.1$. This contains 36 strong MgII absorbers with $EW_0 > 0.3$ Å with redshift up to 1.944 (26 sightlines containing single MgII systems and five sightlines containing two MgII systems).

Figures 1 and 2 clearly show that the distribution of RM for sightlines that pass through a strong MgII system is significantly broader than that for quasars in which absorption is absent. A two-sided Kolmogorov-Smirnov (KS) test rejects the possibility that the distributions with $N_{MgII}=0$, $N_{MgII}> 0$ ($N_{MgII}=0$, $N_{MgII}=2$) come from the same parent distribution at the 92.2 (99.93) percent level, or 96.1 % for the sightlines with quasar redshift $z > 1$. It should be noted that the greatest deviation occurs at RM ~ 40 rad/m$^2$ which corresponds to the median RM of the sample, showing that the result is produced by the bulk of the data and not by a few extreme outliers. This result is obtained without any subtraction of a Galactic foreground model, but remains essentially unchanged if the individual foreground corrections as in reference [6] are applied. Statistical completeness and/or sensitivity to the Galactic foreground subtraction have been major challenges in previous analyses[16].
Since the probability of intersecting an MgII absorber along the sightline increases strongly with the redshift of the quasar, one might worry that the correlation of |RM| with MgII absorption systems in Figures 1 and 2 could arise because of an underlying correlation

between the intrinsic component of the quasar RM and the quasar redshift, e.g. due to an evolutionary effect in the quasars themselves. However, the correlation between |RM| and the redshift of the quasar $z$ is evidently much weaker than that between |RM| and $N_{MgII}$: the former has Kendalls' $\tau=0.11$ and a chance probability of 0.18, while the latter has $\tau=0.25$ and the chance probability of only 0.008. A more transparent test is to simply compare the |RM| of those quasars that have MgII absorbers with the median |RM| of the seven quasars without absorbers that lie closest in redshift. If MgII absorption is not contributing to the RM, then the |RM| of absorber quasars should be randomly distributed above and below this median. We find 20 of the 31 absorber quasar |RM| lie above the local median of non-absorber quasar |RM|, a result which has a probability, from simple binomial statistics, of occurring by chance of 7%. We conclude that the observed correlation between |RM| and $N_{MgII}$ is a real physical effect.

The MgII absorption systems are known to occur in the halos of normal galaxies[7,8] characterized by a wide range of colors and optical luminosities (0.1-3.0 L*), with a covering factor close to unity for impact parameters $D \leq 50$ kpc[9]. This means that it is possible to find a parent galaxy for virtually all MgII systems within this radius, and that one finds few sightlines that pass this close to a galaxy that do not produce MgII absorption. More recent investigations detected MgII absorption even out to distances as large as ~ 100 kpc[10,11] but with a smaller covering factor[11].

With the new information about the individual sightlines from the optical spectra, we can now obtain an improved determination of the RM contribution of the intervening systems with respect to that in reference[6] (see SI). We replace the statistical description of the occurence of MgII systems with the actual information for each sightline. We keep the intrinsic Lorentzian component to a constant (rest-frame) $\Gamma_{intr}/2 = 7^{+6}_{-2}$ rad/m² as obtained in our previous analysis[6], because the many sources at low redshifts $z < 0.6$ in that work allowed an optimal determination of this parameter. Our new calculation yields

$\sigma_{MgII} = 140^{+80}_{-50}$ rad/m² and $\sigma_{noise} = 20^{+10}_{-4}$ rad/m² ($\pm 1\sigma$ uncertainties) for the dispersion of the rest frame RM distribution of MgII systems and Galactic noise, respectively. At redshift $z \sim 0.9$, this leads to an additional RM component in the observer's frame of about 40 rad/m², as seen in Figs. 1 and 2. These numbers suggest that the largest contribution to the observed Faraday Rotation of typical quasars may well be coming from the intervening galaxies, although undoubtedly some individual quasars will exhibit high intrinsic RM[18].

The value of $\sigma_{MgII}$ obtained above for the intervenors is intriguing as it is comparable to the dispersion characterizing the RM data of known galactic systems at low redshift; for example, the disk of the Milky Way[19], with |RM| up to a few hundred rad/m², the Large Magellanic Cloud[20] and a spiral galaxy with an absorbing system at z=0.395 in front of the quasar PKS 1229-021 (ref. 21).

For an order of magnitude estimate of the magnetic field strength, $B$, in the intervening systems, we can re-write the RM definition (see SI) in the rest frame of the MgII system as

$RM(rad/m^2) \simeq 1.5 \cdot 10^{-19} N_e(cm^{-2}) B(\mu G)$, where $N_e$ is the column density of free electrons. Typically MgII systems with $0.3$ Å $<$ EW$_r$ $< 0.6$ Å are found to be associated with galaxies with a column density of neutral hydrogen of $N(HI) \approx 10^{19} cm^{-2}$ [22] and a hydrogen ionization fraction of $\bar{x} \simeq 0.90$ [23,24], yielding $N_e \approx \bar{x}/(1-\bar{x}) N(HI)$, so that

$$B = 10 \left(\frac{\sigma_{MgII}}{140\, rad/m^2}\right) \left(\frac{N_e}{9 \cdot 10^{19} cm^{-2}}\right)^{-1} \mu G.$$

This order of magnitude estimate ignores any field reversals within the system which would boost the required magnetic fields by the square root of the number of such reversals. The optical and radio sightlines to quasars showing MgII absorption are aligned within 2 arcsec (25 kpc at $z = 1$) in over 90 percent of the objects. Most of the radio sources are also compact with angular sizes of about 1 arcsec, i.e. covering an area of less than 10 kpc in the redshift range of $0.6 < z < 2.0$. Since the impact parameters of MgII systems generally extend out to 50 kpc[9], or even to 100 kpc[6,10], from the parent galaxies, it is thus likely that our radio sightlines are probing the outer halos of galaxies in addition to the inner disks. The magnetized, enriched material could have been ejected by galactic winds. However, further observations to locate the parent galaxies of these particular MgII systems relative to the individual radio sightlines will be required before we can fix the location of the magnetized material within these galaxies.

We conclude that magnetic fields in and around completely normal galaxies were, at the time when the Universe was only about one-third (z~1.3) of its present age, already of comparable strength to those that are observed in today's galaxies[25,26]. This reduces significantly the number of e-foldings available for the build up of coherent magnetic fields for alpha-omega dynamo mechanisms in galaxies, setting new strong general constraints for the efficiency of those models[3,27]. It also serves as a general reminder of the potential importance of magnetic fields, usually completely ignored, in the formation and evolution of cosmic structures in the high redshift Universe.

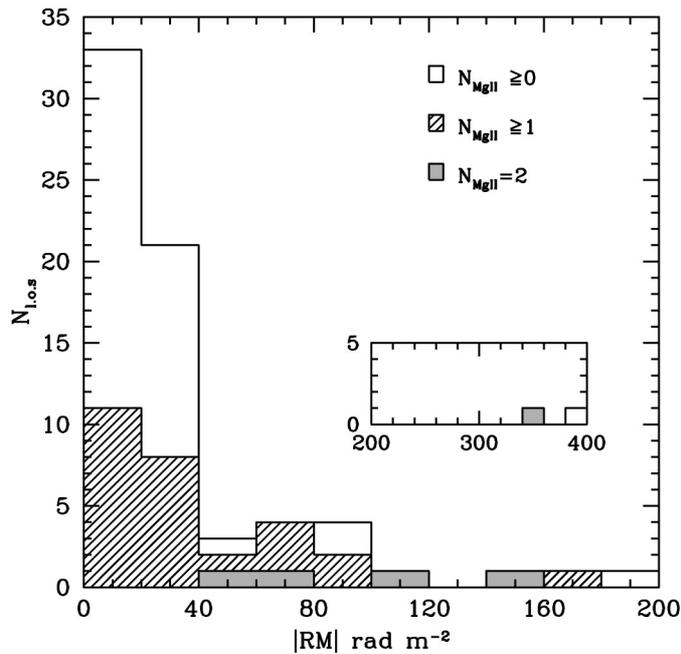

**Figure 1: Distributions of RM for different numbers of strong MgII absorption lines.** The RM histogram of the total sample consisting of 71 sightlines to quasars with redshifts in the range 0.6-2.0 is shown in the background. Overplotted is the hatched histogram of 31 sightlines with one or two strong MgII absorption line systems. The 5 sightlines with two absorption line systems are marked by the grey shaded area. Using Kendall's τ statistic, we determine that |RM| and the number of detected MgII systems are correlated at a significance level of 99.12 percent. A typical RM measurement error is around 3 rad m$^{-2}$.

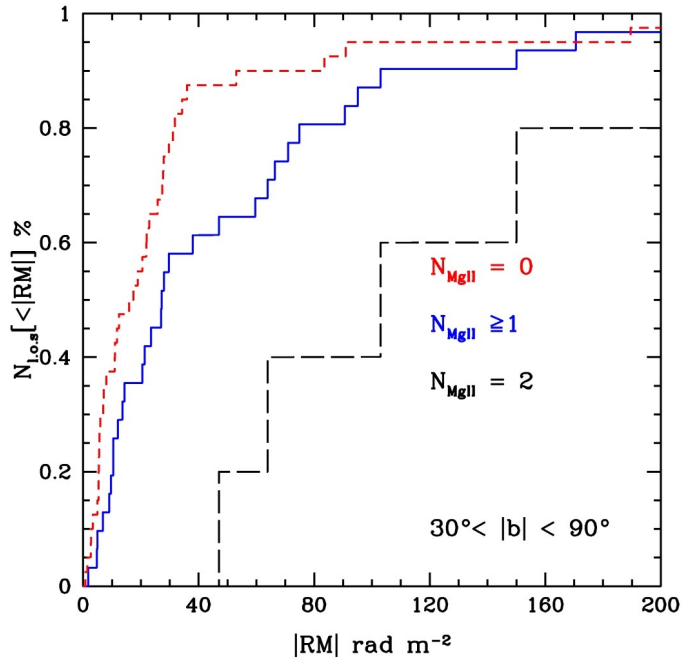

**Figure 2: Cumulative distributions of Rotation Measures of sightlines with and without strong MgII absorption line systems.** The 31 sightlines with one or two strong MgII systems correspond to the blue solid line, those with two strong MgII systems to the black dashed line. The 40 sightlines without strong MgII absorption lines are represented by the red dashed line. For clarity, only the range 0 < RM < 200 rad/m² is shown. A typical RM measurement error is around 3 rad m$^{-2}$. A KS-test indicates that the distributions of RM for $N_{MgII}$=0 and $N_{MgII}$>0 are different at the 92.2 % (96.1% for $z_{qso}$ > 1) and those for $N_{MgII}$=0 and $N_{MgII}$=2 at the 99.93 % significance level.